\documentclass[preprint,aip,rsi]{revtex4-1}
 
\usepackage{graphicx}
\usepackage{amsmath,amssymb}
\usepackage{natbib}

\begin{document}

\title{Eliminating the broadening by finite aperture in Brillouin spectroscopy}

\author{R. Vialla}
\author{B. Ruffl\'e}\email{Benoit.Ruffle@univ-montp2.fr}
\author{G. Guimbreti\`ere}\altaffiliation{now at CNRS CEMHTI, UPR 3079, F-45071 Orleans, France}
\author{R. Vacher}
\affiliation{Universit\'e Montpellier 2, Laboratoire Charles Coulomb UMR 5221, F-34095, Montpellier, France}
\affiliation{CNRS, Laboratoire Charles Coulomb UMR 5221, F-34095, Montpellier, France}

\date{\today}

\begin{abstract}
We present a new optical arrangement which allows to avoid the broadening by finite aperture in Brillouin spectroscopy. In this system, all the rays scattered at the same angle by the whole scattering volume are collected on a single pixel of the area detector. This allows to use large collection angles, increasing the luminosity without lowering the accuracy of the frequency-shift and linewidth measurements. Several results of experimental checks are provided, showing the efficiency of the device. 
\end{abstract}

\maketitle

\section{Introduction}

Brillouin scattering originates from the interaction of light with thermal acoustic modes in matter \cite{Bri22}. It is often used for the measurements of the velocity and attenuation of these modes in solids and in liquids. In an isotropic medium with a refractive index $n$, the interaction of a monochromatic light beam of angular frequency $\omega_0$ with one acoustic mode of velocity $v$ produces a doublet shifted from $\omega_0$ by $\pm \delta \omega_{\rm B} = \pm \frac{2 n v \omega_0}{c} \sin \frac{\theta}{2}$, where $c$ is the velocity of light in vacuum and $\theta$ is the scattering angle. The fullwidth at half maximum of the Brillouin line is related to the attenuation $\alpha_s$ of the mode by $\Gamma = \alpha_s v$. The common values for the frequency shift $\delta \omega_{\rm B} / 2\pi$ range from 1 to 50 GHz, with linewidths $\Gamma / 2\pi$ usually between 1 MHz and 1 GHz. Thus a resolving power $\mathcal{R}$ larger than 10$^6$ is required for the measurement of frequency shifts, while the investigation of linewidths often  imposes $\mathcal{R} \simeq 10^8$. In those ranges, Fabry-Perot (FP) interferometers must be used. In the most common instruments, called spectrometers, the light is collected on a single-channel detector, either a photomultiplier or an avalanche photo-diode, and the spectrum is explored by scanning the optical thickness of the FP \cite{Duf48,*Jac60}. The use of the spectrometers in Brillouin scattering was proposed by Cecchi \cite{Cec64,*Cec65} and Chiao and Stoicheff \cite{Chi64}. The Brillouin scattering cross-section is weak, so that the luminosity of the resolving instrument must be high, a requirement which must be taken into account in designing the FP interferometer. The advent of very sensitive area detectors offers a new possibility. In such instruments, the FP is not scanned but the fixed ring pattern is analyzed \cite{Wal96,Ito96}. This requires in particular the use of larger aperture angles on the FP. Another spectroscopic challenge is the contrast of the resolving system. The above orders of magnitude show that the weak Brillouin doublet at the frequency $\omega_0 \pm \delta \omega_{\rm B}$ has to be studied in the tail of the elastic scattering at frequency $\omega_0$. The intensity of the latter, often arising from unavoidable parasitic light, is usually several orders of magnitude larger than that of the Brillouin line. The contrast of a single-pass FP with reasonable values of the reflection coefficient of the coatings is limited to around 10$^3$, which often implies the use of a multi-pass configuration \cite{San71}.

A specific problem in Brillouin scattering originates from the aperture of the scattered beam. From the above expression, it is seen that $\delta \omega_{\rm B}$ depends on the scattering angle $\theta$, meaning that the light scattered in each elementary solid angle changes in frequency. When the light emitted by the whole aperture angle is collected on a single detector, this produces an artificial broadening, and thus an uncertainty on the measurement of both $\delta \omega_{\rm B}$ and $\Gamma$. This problem is particularly important for the measurement of small linewidths or when small scattering volumes and thus large apertures are used, such as in confined media ({\it e.g.} optical fibers or thin films) or in micro-Brillouin set-ups. In this paper, we present a new optical arrangement which provides an efficient solution to this problem.

\section{Principle of the Brillouin spectrograph}

\begin{figure}
\includegraphics[width=7cm]{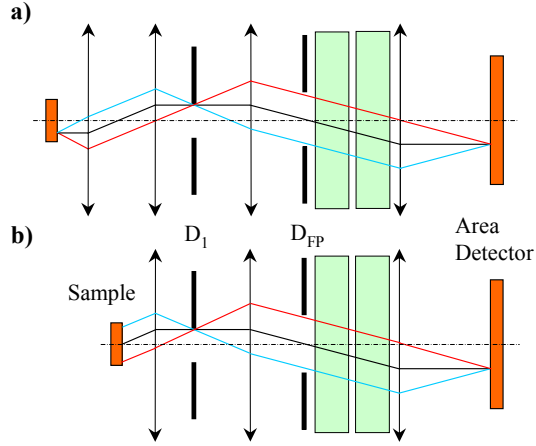}
\caption{\label{Fig1} (Color online) a) Typical optical arrangement used in Brillouin spectrometers. The scattering volume is imaged on the detector. b) Optical arrangement proposed in our Brillouin spectrograph. The light scattered in a given direction by the whole scattering volume is focused on one channel of the detector.}
\end{figure}

In the following, the instrument using an area detector is called a spectrograph. A typical optical arrangement used in Brillouin scattering for such instruments is shown in Fig. \ref{Fig1}a. In this configuration the scattering volume is imaged on the sensitive surface of the detector. In this case the light scattered in the whole aperture angle by an elementary volume of the sample is collected on one channel of the detector, leading to the artificial broadening described above. In our arrangement, shown in Fig. \ref{Fig1}b, the light scattered in a given direction by the whole scattering volume, {\em i.e.} a parallel beam defined by polar coordinates ($i, \varphi$) is focused on one channel of the detector. Therefore, this single channel receives Brillouin scattering emitted with a single scattering angle $\theta$. Of course, $\theta$ now changes from one channel to the other one, but the integration over the variation of $\theta$ in the scattering aperture is avoided. The variation of $\theta$ between channels can easily be taken into account in the data treatment. In order to ensure the quality of the analysis, the etendue in which the scattered light is collected must be defined precisely. In the arrangement shown in Fig. \ref{Fig1}b, the sample is placed at the focus of the first lens. A diaphragm D$_1$ at the second focus of this lens limits the solid angle emitted by each scattering volume. By this arrangement, each point in the scattering volume scatters in the same solid angle. The diaphragm D$_1$ is imaged on the area detector. On the other hand, the scattering volume is imaged on the surface used of the FP, limited by a diaphragm D$_{\rm FP}$. The diaphragms D$_1$ and D$_{\rm FP}$ define the scattering etendue. It is also important for the data treatment to know precisely the polar coordinates ($i, \varphi$) of the light collected on each individual detector. This imposes to minimize geometrical aberrations.

It has been shown that, in high resolution Brillouin scattering measurements, a confocal FP \cite{Con58} must be used \cite{Pin68,Pel75}. However, with the high contrast needed for most experiments, the confocal FP which cannot be used practically in multi-pass must be combined with a plane FP in a tandem configuration. For simplicity, we concentrate here to a first spectrograph configuration using a multi-pass plane FP. In fact, it has been shown that for high contrast, the multi-pass configuration has a much higher luminosity than an increase of the reflection coefficient in a single-pass FP \cite{San71}.

With standard 2-inches mirrors in a multi-pass configuration, the surface used in each pass is limited practically to a circle of 10 mm diameter. On the other hand, if we want to cover for instance a spectral domain of 10 GHz by the angular dispersion of the FP, a half-aperture angle of nearly 6 mrad must be allowed. The corresponding etendue is $U = 9 \times 10^{-9}$ m$^2$, which must be collected from the scattering volume. When this etendue is imaged with that scattered by the sample, it must be kept in mind that the flux collected at constant etendue of the spectrograph increases linearly with the magnification $\gamma$ of the whole optical collection system, thus encouraging the use of high $\gamma$. If we take for example $\gamma=50$, the diameter of the scattering surface is $d=0.2$ mm and the half scattering angle is 0.3 rad. With such large apertures, the first collecting lens in the optical arrangement will not be working in the Gauss approximation.

We must also consider that when the plane FP is used with large incidence angles, the so-called "walk-off" effect must be taken into account \cite{Wal96}: each double reflection of an optical ray with an angle $\alpha$ inside the FP of thickness $e$ induces a lateral shift equal to $2e\alpha$. For large values of $\alpha$, the number of multiple reflections allowed can be limited by the geometry of the FP arrangement. Practically, the transmission function is no longer an Airy profile, but a function somehow broader which must be calculated for each value of $\alpha$ in the data treatment.

\section{Description of the instrument}

\begin{figure}
\includegraphics[width=8.5cm]{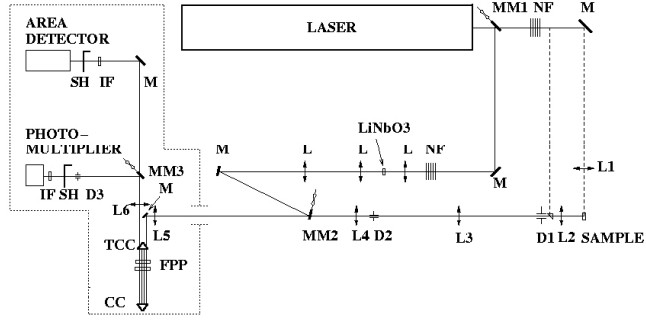}
\caption{\label{Fig2} Schematic representation of the Brillouin spectrograph. L: lens; MM: Mobile Mirror; M: Mirror; SH: Shutter; D: Diaphragm; NF: Neutral Filter; TCC: Truncated Corner Cube; CC: Corner Cube; IF: Interference Filter.}
\end{figure}

A schematic representation of our device is shown in Fig. \ref{Fig2}. The source is a single-frequency argon-ion laser. We use the line at 514.5 nm. In the experiments described below, the power was equal to 200 mW. The beam is focused on the sample in order to reduce the diameter of the incident beam to the appropriate value. Two possible scattering geometries, $\theta = 90^\circ$ and $\theta = 180^\circ$, are shown. 

We choose for the collecting system a magnification $\gamma \simeq 50$. This is realized practically by two telescopic arrangements. The sample is located at the focus of the first lens L$_2$. The first telescopic system comprises the lenses L$_2$, a Gradium lens with a focal length $f_2$ = 125 mm and L$_3$, with $f_3$ = 500 mm. The value of $f_2$ must be large enough to allow the use of standard sample environments. The second telescopic system is made of lenses L$_4$, a second Gradium lens ($f_4$ = 40 mm), and L$_5$ ($f_5$ = 500 mm). The focuses of L$_3$ and L$_4$ coincide on the diaphragm D$_2$. In practice, D$_1$ and D$_2$ define the scattering etendue. All the lenses in the instrument are plane-convex. Finally, the area detector is located at the focus of the lens L$_6$. In this way, both the ring figure of the FP and the diaphragm D$_1$ are imaged on the detector surface. In order to minimize the divergence of the beam across the 4-pass FP, the image of D$_2$ is placed between the second and the third pass. 

The area detector is an Andor DV434 CCD with 1024$\times$1024 pixels cooled by a 4-stage Peltier refrigerator at a temperature of -70$^\circ$C. The dark count is about  6e$^-$/pixel/hour. The total area is around 13 $\times$ 13 mm$^2$. In order to cover reasonably this area we take $f_6$ = 1000 mm. 

We have chosen to use a 4-pass FP in order to have a contrast large enough. On the other hand, we want to be able to adapt the resolving power by changing the distance $e$ between plates. These two constraints forbid the use of a solid etalon or of a fixed spacer between plates. The appropriate configuration in this case is that proposed by Sandercock \cite{San71}. In this system, the scan is realized by displacement of one of the plates without loss of parallelism. The orientation of one of the plates can be adjusted to optimize this parallelism. The drawback of such a system is that it must be continuously adjusted, which impose to realize multiple storing periods separated by an optimization procedure. The most important point is that the thickness of the FP must be kept constant during the whole recording of the spectrum. To that purpose, we use an auxiliary frequency-modulated beam in which a line shifted by electro-optic modulation at a frequency $\delta\omega_{\rm M}$ near $\delta\omega_{\rm B}$ is produced \cite{Sus79}. This "Regulation" beam is also shown in Fig. \ref{Fig2}. A set of mobile mirrors, MM$_1$, MM$_2$ and MM$_3$ directs the incident beam in the electro-optic cell, a resonant cavity containing a LiNbO$_3$ crystal excited by an electromagnetic microwave generator. The modulated light is then directed to the FP in normal incidence, and finally to a photomultiplier. The feedback system optimizes the transmission of the modulated light at frequency $\omega_0 + \delta\omega_{\rm M}$ or $\omega_0 - \delta\omega_{\rm M}$, by acting on the thickness $e$ and on the orientation of one of the plates. The frequency of this modulated light is transmitted in normal incidence by the FP, defining a frequency origin for the ring pattern. Thus the choice of the value of the difference $\delta\omega_{\rm B} - \delta\omega_{\rm M}$ will fix the radius of the ring on the area detector to the appropriate value. The ring pattern is then calibrated in frequency when the distance between L$_6$ and the area detector is known. Presently, we are working on another regulation system in which the photomultiplier is replaced by an area detector. In this case we choose 
$\delta\omega_{\rm B} = \delta\omega_{\rm M}$, and the adjustment is monitored by the radius and the intensity of the ring on this auxiliary area detector.

\section{Data analysis, numerical and experimental checks}

The principle of the data treatment consists in considering each parallel beam characterized by $(i,\varphi)$ emitted by the whole scattering volume. The central frequency $\omega_{\rm B}(i,\varphi)$ of the Brillouin line is calculated from the sound velocity $v$, taken as a free parameter in the fitting procedure. The linewidth $\Gamma$ of the Brillouin profile is the second fitting parameter. In practice, $\omega_{\rm B}(i,\varphi)$ changes by about 10 \% over the whole detected Brillouin annulus, and thus the frequency dependence of $\Gamma$ can be neglected in regard of the resolution of the instrument. In our 4-pass arrangement, this parallel beam reaches the diaphragm D$_{\rm FP}$ located between the second and the third passes under angles ($\alpha$, $\beta$) which must be calculated. Those angles define the transmission of the 4-pass FP for each frequency $\omega$. At this stage, the walk-off effect, depending not only on ($\alpha$, $\beta$) but also on the coordinates ($x, y$) at which each elementary beam reaches D$_{\rm FP}$, can be taken into account. Integrating over $\omega$, the intensity received by a given pixel of the area detector is obtained.

\begin{figure}
\includegraphics[width=8.5cm]{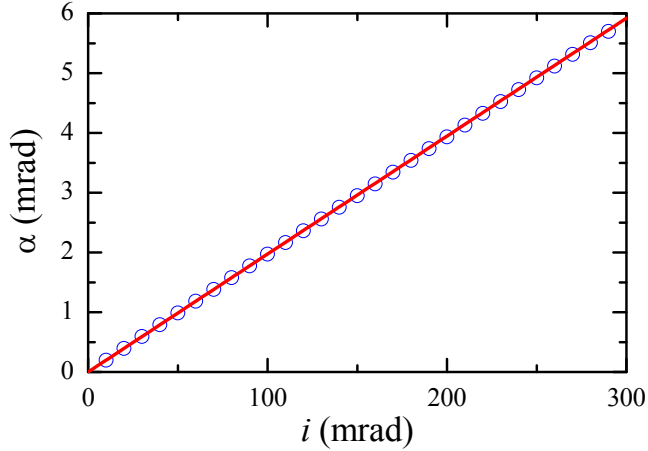}
\caption{\label{Fig3} (Color online) Calculated incident angle $\alpha$ on the FP interferometer as a function of the angle $i$ of a beam emitted by the scattering volume (circles). The line $\alpha = i/\gamma$ is the expected behavior for an aberration-free optical system with $\gamma$ = 50.67.}
\end{figure}

First, we have used a model of the optics of our device to check the quality of the correction for aberrations. We have tested the relation between the angle $i$ at the scattering volume assumed to be a point on the optical axis, and the angle $\alpha$ for the same ray impinging the FP interferometer. Of course, for a perfect optical system in the Gauss approximation, $i = \gamma\alpha$. The calculated $\alpha$ {\it vs} $i$ is plotted in Fig. \ref{Fig3}. The result is very close to the perfect straight line. The departure, reaching only 0.4 \% at the largest angles, can be taken into account in the data treatment.

\begin{figure}
\includegraphics[width=8.5cm]{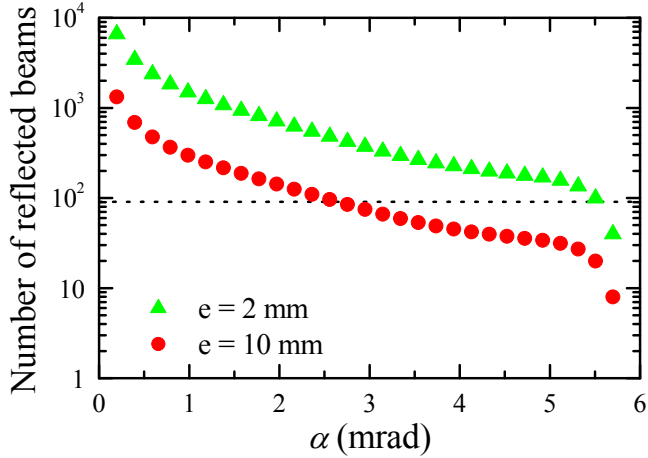}
\caption{\label{Fig4} (Color online) Number of reflected beams in the FP interferometer {\it vs} incident angle $\alpha$ for two thicknesses $e$. The dashed line at $2 \mathcal{F}_1\simeq 90$ sets the minimum number of interfering beams to maintain both the finesse and the transmission.}
\end{figure}

Another important check is the influence of the walk-off on the transmission of each elementary beam. As an example, we have calculated the number of reflected beams as a function of $\alpha$ for the beam emitted by the geometrical center of the scattering volume, reaching a FP interferometer limited by a diaphragm of 10 mm diameter. The results are plotted in Fig. \ref{Fig4} for two different thicknesses, 2 and 10 mm. It shows that the number of reflected beams is large enough to keep the 1-pass finesse $\mathcal{F}_1 \simeq 45$ of the FP over the whole aperture for $e$ = 2 mm, and up to angles $\alpha \simeq$ 3 mrad for $e$ = 10 mm. In practice, the effect of the walk-off can be fully calculated, taking into account the geometry of the 4-pass FP interferometer.

\section{experimental results}

\begin{figure}
\includegraphics[width=8.5cm]{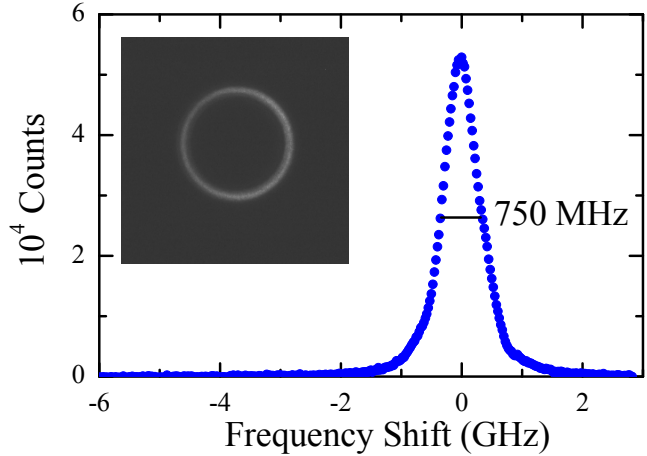}
\caption{\label{Fig5} (Color online) Instrumental profile for $e$ = 2 mm, showing a FWHM of 750 MHz. (inset) Ring pattern of the elastic scattering light of liquid glycerol.}
\end{figure}

In order to determine experimentally the instrumental function of the spectrograph, we have recorded the elastic scattering line of a liquid glycerol sample at room temperature. The obtained ring pattern is shown in the inset of Fig. \ref{Fig5}. The profile deduced from the analysis of this ring has a full-width at half-maximum (FWHM) equal to about 750 $\pm$ 10 MHz, which corresponds to a finesse $\mathcal{F}\simeq$ 100. This is in excellent agreement with the value calculated for a 4-pass FP with a reflectivity $R$ = 0.94 and a flatness of $\lambda/200$. The finesse obtained with a spectrograph using an area detector is higher than that of the same FP in a scanning configuration, as the broadening of the transmission function by the aperture supported by the pinhole is avoided.

To demonstrate the efficiency of the spectrograph, we have performed experiments in the right-angle configuration, rather than in backscattering, as  it is well-known that the parasitic effect due to the angular aperture of the scattered beam is minimum in the latter. Further, transverse modes are silent in backscattering for isotropic media.

\begin{figure}
\includegraphics[width=8.5cm]{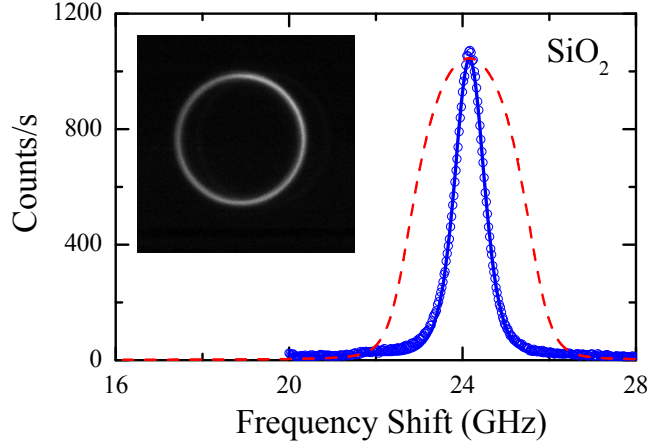}
\caption{\label{Fig6} (Color online) Brillouin spectrum from longitudinal modes of vitreous silica in the right-angle geometry derived from the CCD image. The solid line is a fit of the ring pattern to a damped harmonic oscillator convoluted with the transmission function of the FP. The dashed curve is a calculation of the Brillouin spectrum which would have been obtained with a scanning spectrometer as described in the text. (inset) Ring pattern after an exposure time of 20 seconds.}
\end{figure}

Fig. \ref{Fig6} shows the results for Brillouin scattering by longitudinal modes in vitreous silica, with the ring pattern in the inset. The spectrum was recorded during 20 seconds. After having processed the ring pattern as explained in the previous section, the Brillouin spectrum at 90$^\circ$ is derived by an azimuthal integration, taking into account the variation of $\omega_{\rm B}$ with $(i,\varphi)$. The area of the curve represents the total number of counts received by the detector during this period. An excellent signal-to-noise ratio is obtained in this short exposure time. On the other hand, the Brillouin profile is very close to the instrumental line, showing that the aperture broadening has been totally avoided. The dashed line is a calculation of the spectrum which would have been recorded by a spectrometer working with the aperture corresponding to the average radius of the ring pattern. It clearly emphasizes the advantage of our optical arrangement. The solid line is the proper azimuthal integration of a damped harmonic oscillator (DHO) convoluted with the transmission of the FP adjusted to the ring pattern. From the fit of the Brillouin ring, we measure a Brillouin frequency shift, $\omega_{\rm B}/2\pi = 24.16 \pm 0.01$ GHz in agreement with previous measurements \cite{Pel75}. The obtained FWHM $\Gamma/2\pi$ = 120 $\pm$ 20 MHz can be compared with the result $\Gamma \simeq 100 \pm$ 6 MHz found with a high-resolution spectrometer \cite{Vac06}. The agreement is satisfactory, regarding the low resolution of our plane FP instrument.

\begin{figure}
\includegraphics[width=8.5cm]{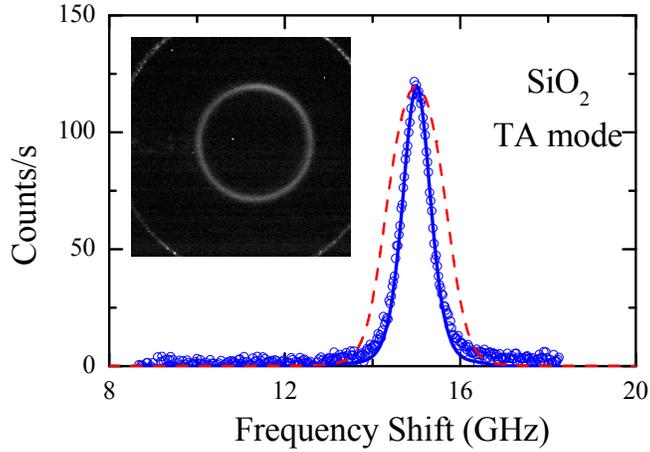}
\caption{\label{Fig7} (Color online) Brillouin spectrum from transverse modes of vitreous silica in the right-angle geometry. (inset) Ring pattern after an exposure time of 40 seconds.}
\end{figure}

To demonstrate the high sensitivity of the spectrograph, we have recorded Brillouin scattering from transverse acoustic modes in vitreous silica. The spectrum and the ring pattern obtained in 40 seconds are shown in Fig. \ref{Fig7}. Not only the profile is very well defined after this short exposure time, but also the contrast is very high. The Brillouin line is observed on a flat background, thus perfectly separated from the wings of the elastic line. This is a clear advantage of the multi-pass configuration of our instrument. As in Fig. \ref{Fig6}, the dashed line shows the spectrum which would have been obtained with a spectrometer working with the same collection angle. The FWHM of the transverse Brillouin line at 90$^\circ$ and at room temperature is known to be about 40 MHz \cite{Vac75,Ruf10}, a value too small to be measured in the present configuration of the instrument. The solid line in Fig. \ref{Fig7} is thus the calculation of the profile obtained with a DHO Brillouin profile having a FWHM of 40 MHz. It is very close to the instrumental line. We measure $\omega_{\rm B}/2\pi = 15.00 \pm 0.03$ GHz in agreement with previous measurements \cite{Vac75}.

\begin{figure}
\includegraphics[width=8.5cm]{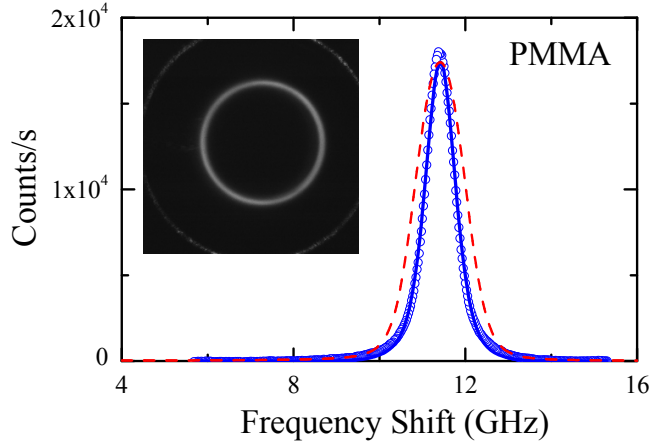}
\caption{\label{Fig8} (Color online) Brillouin spectrum from longitudinal modes of amorphous PMMA in the right-angle geometry. (inset) Ring pattern after an exposure time of 4 seconds.}
\end{figure}

\begin{figure}
\includegraphics[width=8.5cm]{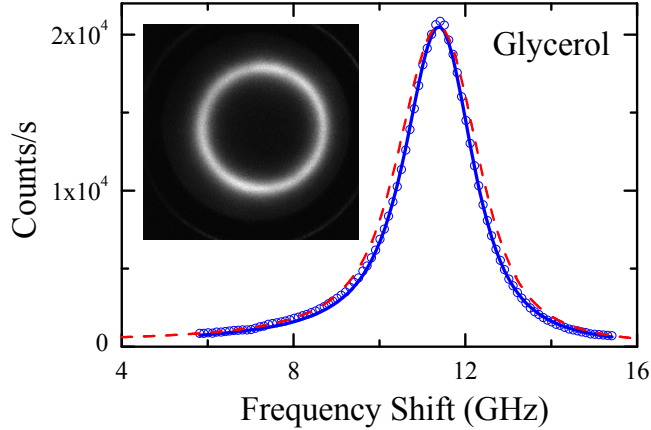}
\caption{\label{Fig9} (Color online) Brillouin spectrum from longitudinal modes of liquid glycerol in the right-angle geometry. (inset) Ring pattern after an exposure time of 4 seconds.}
\end{figure}

Other examples of Brillouin spectra of longitudinal acoustic modes at room temperature are shown in Fig. \ref{Fig8} and \ref{Fig9} for amorphous Poly(methyl methacrylate) (PMMA) and liquid glycerol, respectively. Both were obtained in 4 seconds. For PMMA, we measure $\omega_{\rm B}/2\pi = 11.42 \pm 0.01$ GHz in excellent agreement with a previous determination \cite{Jac72}. We find $\Gamma/2\pi = 160 \pm 15$ MHz. When compared to the value $\Gamma/2\pi = 200 \pm 10$ MHz known from backscattering measurements \cite{Vac76} obtained with a high resolution spectrometer, this indicates a frequency dependence $\Gamma \widetilde{\propto} \Omega^{0.6}$ in this temperature-frequency domain. In the case of glycerol, the FWHM is large enough that the broadening by integration over the finite aperture has little influence on the profile, as seen in Fig. \ref{Fig9}. The measured values are $\omega_{\rm B}/2\pi = 11.45 \pm 0.03$ GHz and $\Gamma/2\pi = 1.52 \pm 0.02$ GHz which compare very well with values reported in a recent study of the dynamic structure factor of glycerol in the backscattering geometry using a Sandercock-type FP-interferometer \cite{Com03}.

\section{Conclusion}

Using an original optical arrangement, we have realized a Brillouin spectrograph in which the broadening by finite aperture is fully eliminated. This instrument was successfully used with the scattered beam having an aperture as large as 0.3 rad. The 4-pass configuration for the plane FP provides contrast high enough that weak signals such as transverse Brillouin lines are clearly measured without parasitic elastic scattering background. The finesse is found equal to 100 as expected. The luminosity of the spectrograph is very high, as already emphasized by several authors using area detectors \cite{Wal96,Ito96}. As all spectral intervals are recorded simultaneously rather than scanned, the power received on the detector is roughly multiplied by a factor equal to the finesse.

This instrument will allow new developments in applications of Brillouin scattering which were difficult up to now. The accuracy of frequency measurements in experiments were large collecting apertures must be used, such as micro-Brillouin and investigations of small scattering volumes, will be greatly enhanced. Both the increased accuracy and the short exposure times will allow investigation of rapidly changing phenomena such as relaxation in glass-forming liquids near the glass transition temperature. Experiments at various scattering angles allowing frequency variations and studies of weak transverse acoustic modes will be greatly facilitated.

\section{Acknowledgments}

It is a pleasure to thank P. Solignac, P. Martinez, E. Arnould, and J.M. Fromental from the Laboratoire Charles Coulomb, Montpellier, as well as undergraduate students from Lyc\'ee Champollion, Montpellier, France, and from ISMRA, Caen, France, for technical developments, and to Dr. P.  Falgayrettes from the Institut d'Electronique du Sud, Montpellier for fruitful discussions. This work is partially funded by the Agence Nationale pour la Recherche (Grant No. ANR-08-MATPRO-0473-85, {\em Postre}) and R\'egion Languedoc-Roussillon (Omega Platform).


%

\end{document}